# Growth Conditions and Interfacial Misfit Array in SnTe (111) films Grown on InP (111)A Substrates by Molecular Beam Epitaxy

Qihua Zhang[1], Maria Hilse[1,2,3], Wesley Auker[2], Jennifer Gray[2], Stephanie Law[1,2,3,4]

[1] Two Dimensional Crystal Consortium Materials Innovation Platform, The Pennsylvania State University, University Park, PA 16802, USA

[2] Materials Research Institute, The Pennsylvania State University, University Park, Pennsylvania 16802 USA

[3] Department of Materials Science and Engineering, The Pennsylvania State University, University Park, Pennsylvania 16802 USA

[4] Institute of Energy and the Environment, The Pennsylvania State University, University Park, PA 16802 USA

ABSTRACT Tin telluride (SnTe) is an IV-VI semiconductor with a topological crystalline insulator band structure, high thermoelectric performance, and in-plane ferroelectricity. Despite its many applications, there has been little work focused on understanding the growth mechanisms of SnTe thin films. In this manuscript, we investigate the molecular beam epitaxy (MBE) synthesis of SnTe (111) thin films on InP (111)A substrates. We explore the effect of substrate temperature, Te:Sn flux ratio, and growth rate on the film quality. Using a substrate temperature of 340 °C, a Te:Sn flux ratio of 3, and a growth rate of 0.48 Å/s, fully coalesced and single crystalline SnTe (111) epitaxial layers with x-ray rocking curve full-width-at-half-maxima (FWHM) of 0.09° and root-mean-square surface roughness as low as 0.2 nm have been obtained. Despite the 7.5% lattice mismatch between the SnTe (111) film and the InP (111)A substrate, reciprocal space mapping indicates that the 15 nm SnTe layer is fully relaxed. We show that a periodic interfacial misfit (IMF) dislocation array forms at the SnTe/InP heterointerface, where each IMF dislocation is separated by 14 InP lattice sites/13 SnTe lattice sites, providing rapid strain relaxation and yielding the high quality SnTe layer. This is the first report of an IMF array forming in a rock-salt on zinc-blende material system and at an IV-VI on III-V heterointerface, and highlights the potential for SnTe as a buffer layer for epitaxial telluride film growth. This work represents an important milestone in enabling the heterointegration between IV-VI and III-V semiconductors to create multifunctional devices.

## 1. Introduction

Tin telluride (SnTe) is a narrow bandgap semiconductor with many desirable properties: it has a high thermoelectric figure of merit with nontoxic components[1-7], it is a topological crystalline insulator[8-14], and it exhibits ferroelectric behavior when its crystal structure undergoes a phase transition.[15-19] Bulk SnTe crystals have a structural phase transition temperature ($T_C$) of ~100 K,[15] but $T_C$ can be greatly enhanced when the thickness of a SnTe thin film is reduced. SnTe films with a thickness of 1 unit cell (6.32 Å) have shown a $T_C$ of 270 K.[20-24] The discovery of near-room-temperature $T_C$ makes SnTe thin films feasible for a wide range of ferroelectric applications including non-volatile memory devices and non-linear optoelectronics.[18, 20, 24-26] In addition, the band structure of SnTe is topologically non-trivial. Theoretical calculations predict that gapless surface states can be observed on both the (001) and (111) planes of SnTe, due to the mirror symmetry of its rock-salt crystal structure.[8, 10] Tanaka *et al.* and Zhang *et al.* were among the first to experimentally confirm the Dirac-cone-like band structure on SnTe (100) and (111) using angle-resolved photoemission spectroscopy (ARPES) measurements.[9, 13]

In addition to its attractive material properties, SnTe also serves as an important buffer layer for Te-based heterostructures. For example, SnTe is a popular back surface field layer for photovoltaic devices due to its narrow bandgap and high hole concentration, and is thus often used as a buffer layer for CdTe-based solar cell heterostructures.[27-29] SnTe can also be used as a spacer material for antiferromagnetic EuTe quantum dots, due to the relatively low lattice mismatch between the two alloys.[30, 31] Very recently, with a thin (2 nm) SnTe buffer layer, (Sn-Pb-In)Te alloys exhibited superconductivity and semimetal behavior with electron mobilities exceeding 5000 $cm^2/Vs$.[32, 33]

The quality of SnTe strongly affects its intrinsic properties as well as the performance of devices incorporating SnTe, so it is critical to understand the growth mechanism of SnTe using a mature synthesis technique. Molecular beam epitaxy (MBE) is a promising technique for the synthesis of high-quality SnTe layers due to its ultra-high vacuum (UHV) growth environment, use of high purity source materials, and *in situ* surface monitoring capabilities. In the past, growths of SnTe films by MBE have been reported by multiple research groups.[33-37] However, few of the efforts have been focused on understanding the growth mechanisms of SnTe films. Masuko *et al.* demonstrated the growth of $Sn_xIn_{1-x}Te$ (111) thin films with a full-width-at-half-maximum (FWHM) of 0.11° in x-ray diffraction (XRD) rocking curves, yet no analysis of surface morphology was provided.[33] Recently, the research group led by M. Kobayashi studied the surface morphology and electronic properties of SnTe (100) layers.[34, 35] However, these films also contained misoriented (111)-SnTe domains. A thorough investigation of how growth conditions impact the quality of SnTe thin films is thus needed.

In addition, to fully realize the potential of SnTe buffer layers, developing a technique for synthesizing ultrathin relaxed layers with a low density of threading dislocations on a lattice-mismatched substrate is crucial. Interfacial misfit (IMF) arrays are a longstanding method for obtaining high-quality relaxed epilayers on heterogeneous substrates. IMF arrays consist of misfit dislocations that are uniformly spaced at the epilayer/substrate heterointerface and which provide rapid and effective strain relaxation (> 99 %) at the interface, thus creating a relaxed epilayer with a low threading dislocation density. In the past, IMF arrays have only been observed in III-V semiconductor epilayers.[38-43] A well-established example is the GaSb epilayer on a GaAs substrate: despite their 7.8 % lattice mismatch, the threading dislocation density of GaSb on GaAs can be on the order of $10^8$ cm$^{-2}$ with the use of an IMF array.[43] Similarly, with the use of an IMF array, AlSb epilayers can be grown with > 98 % strain relaxation on Si (100) substrates, which has a 13% lattice-mismatch to the epilayer.[39] Although there has recently been one report of an IMF array in a (111)-oriented system,[43] there are no previous reports of IMF arrays in a (111)-oriented IV-VI epilayer on a (111)-oriented III-V substrate.

In this work, we use the IMF array technique for the growth of high quality SnTe epitaxial layers on InP (111)A substrates, which are chosen for their low cost and large-scale availability. The substrate has a 7.5 % lattice mismatch to SnTe,[44] similar to that in the GaSb/GaAs material system. A downside of using an InP (111)A substrates is the difference in the crystal structure between SnTe (rock-salt) and InP (zinc-blende). Despite these differences, we identify an optimized growth window comprising a substrate temperature in the range of 300-340 °C, a Te:Sn flux ratio of ~ 3, and growth rate of 0.48 Å/s, which yields fully-coalesced, single crystal, relaxed SnTe (111) films with a FWHM less than 0.1° in the SnTe (222) XRD rocking curve and a root-mean-square (RMS) film surface roughness as low as 0.2 nm in atomic force microscopy (AFM). From scanning transmission electron microscopy (STEM) analysis at the SnTe/InP heterointerface, we find an IMF dislocation array, where each misfit dislocation is separated evenly by 14 InP lattice sites and 13 SnTe lattice sites, which relieves the strain and yields the high quality SnTe epilayer. This result marks the first demonstration of an IMF array in the rock-salt/zinc-blende system, as well as the integration of IV-VI films on III-V semiconductors.

## 2. Experiments

As-received 2-inch undoped InP (111)A wafers (WaferTech) were first diced into $1 \times 1$ cm$^2$ pieces. The diced substrate was then submerged sequentially in acetone and isopropyl alcohol in an ultrasonic bath to remove surface contamination, followed by a DI water rinse at room temperature. The substrate was further subjected to a UV-ozone treatment for 5 min to degrade organic contaminants. After the cleaning process, the substrate was immediately transferred to the load lock chamber with a base pressure of $5 \times 10^{-9}$ Torr and outgassed at 200 °C for 2 hours before being loaded into the MBE chamber. Prior to growth, the substrate was thermally annealed in the ultrahigh vacuum (UHV) chamber at 600 °C for 10 min to desorb the residual surface oxide. A tellurium flux in the range of $0.28$-$2.3 \times 10^{14}$ cm$^{-2}$/s was supplied during the annealing process. A reflection high electron diffraction (RHEED) system with an electron beam energy of 15 keV (STAIB Instruments and kSA 400 analytical RHEED software) was used to monitor the surface of the substrate as well as the SnTe growth.

All SnTe growths were performed in a DCA Instruments R450 MBE system with a base pressure of $5 \times 10^{-10}$ Torr. The Sn flux was supplied by thermal evaporation of high purity Sn (5N) in a dual-filament effusion cell. The Te flux was supplied by evaporation of Te (5N purity) in a low-temperature effusion cell. Both fluxes were calibrated at room temperature using a ColnaTec quartz crystal microbalance (QCM) operating at 6 MHz using tooling factors determined from physical film thickness measurements prior to SnTe growth. All SnTe layers were grown under Te-rich conditions to minimize Te re-evaporation and suppress metal droplet formation. For the current study, the maximum nominal growth rate of the SnTe film was 0.48 Å/s, where the operating temperature in the Sn effusion cell reached its maximum rated temperature. At the end of growth, all samples were annealed for three minutes in a Te environment at the growth temperature before being cooled to room temperature at a rate of 50 °C/min.

High resolution XRD measurements of the SnTe thin films were performed using a Malvern PANalytical 4-circle X'Pert$^3$ system equipped with a hybrid 2-bounce asymmetric Ge(220) monochromator and PIXcel3D detector with an anti-scatter slit. All XRD scans were taken at room temperature. The 2θ-ω scan was employed to determine the orientation of the as-grown SnTe films. Reciprocal space mapping (RSM) was performed on the SnTe film and the InP substrate on the (222) and (224) diffraction peaks, respectively. X-ray reflectivity (XRR) measurements were used to determine the thickness of the SnTe layers. For XRR, an X-ray mirror (Cu with a Si parabolic mirror and 1/32° divergence slit) was used as the beam incidence optics, along with a parallel plate beam collimator with 0.09 reflectivity slit as the diffracted beam optics. The thicknesses of the SnTe films (Sample A-G, excluding Sample E) were in the range of 12-15 nm. Figure S1 displays an exemplary XRR measurement and the fitting result of Sample G, which has a thickness of ~14.6 nm.

The surface morphology of SnTe layers was examined by a Bruker Dimension Icon atomic force microscope operating in peak force tapping mode with a SCANASYST-AIR probe, which has a nominal spring constant of 0.4 N/m. Scan parameters include a scan rate of 0.5 Hz, a lateral resolution of 512 pixels/line, and a peak force frequency of 2 kHz.

The as-grown SnTe film was prepared for cross-sectional STEM investigation using an FEI Helios Nanolab 660 Dual-Beam Focused Ion Beam (FIB) system. Carbon deposition was achieved using both electron beam and ion beam methods. Initially, electron beam-induced carbon deposition was conducted at 5 kV at 26 nA to create a protective layer on the sample surface. Subsequently, ion beam-induced carbon deposition was performed at 30 kV at 1 nA, enhancing the structural integrity of the deposited layer. For the final thinning and cleaning stages, the ion beam settings were adjusted to a lower energy of 2 kV, optimizing the precision of the milling process and minimizing potential damage from the ion beam. This combination of methodologies allowed for controlled sample manipulation, ensuring high-resolution imaging and analysis. A double aberration corrected FEI Titan3 G2 60-300 S/TEM system operating in STEM mode with a resolution of 0.07 nm was used for acquiring the STEM images of the as-prepared lamella. A high brightness electron source (X-FEG) operating at 300 kV and a high-angle annular dark field (HAADF) detector with a collection angle of 50-100 mrad was used for collecting HAADF-STEM images.

## 3. Results and Discussion

### A. Effect of substrate temperature deduced by *in situ* RHEED

We first studied the effect of substrate temperature ($T_{sub}$) on the growth of SnTe thin films using *in situ* RHEED monitoring. In this experiment, the growth rate was kept at 0.12 Å/s and the Te:Sn flux ratio was maintained at 2. Figure 1(a) shows the RHEED patterns of the InP (111)A substrate along the [11$\bar{2}$] and [1$\bar{1}$0] directions taken after thermal deoxidation. A typical bright, streaky pattern is observed, indicating a clean surface free of In droplets.[45] Figures 1(b-e) subsequently show the RHEED patterns taken immediately after the growth of the SnTe films at $T_{sub}$ of 200 °C, 250 °C, 300 °C, and 350 °C, respectively. At 200 °C, spotty patterns can be observed in both directions with hazy diffraction rods in the [1$\bar{1}$0] direction, indicating the formations of rough 3D islands on the substrate surface. The spacing between the two sets of spotty patterns is consistent with the pattern of the SnTe (001) surface, suggesting that the layer contains a mixture of (111)- and (001)-oriented SnTe domains. As the substrate temperature increases to 250 °C, more pronounced streaks can be observed along the $[11\bar{2}]$ direction, yet spots remain the dominant feature in the $[1\bar{1}0]$ direction. By further increasing the substrate temperature to 300 °C, bright and streaky patterns free of Laue rings and chevrons are seen in both directions (Figure 1(d)), indicating the formation of a single-crystalline SnTe thin film with good in-plane epitaxial alignment. Similar RHEED patterns were recorded for substrate temperatures up to 340 °C (not shown). However, at $T_{sub}$ of 350 °C, solely RHEED patterns similar to the InP (111)A substrate were observed (Figure 1(e)), indicating that no film was deposited due to a high desorption and re-evaporation rate of surface adatoms and Sn-Te nuclei at this temperature. In summary, an optimal window of $T_{sub}$ = 300-340 °C was identified through RHEED for SnTe film growth on InP (111)A substrates.

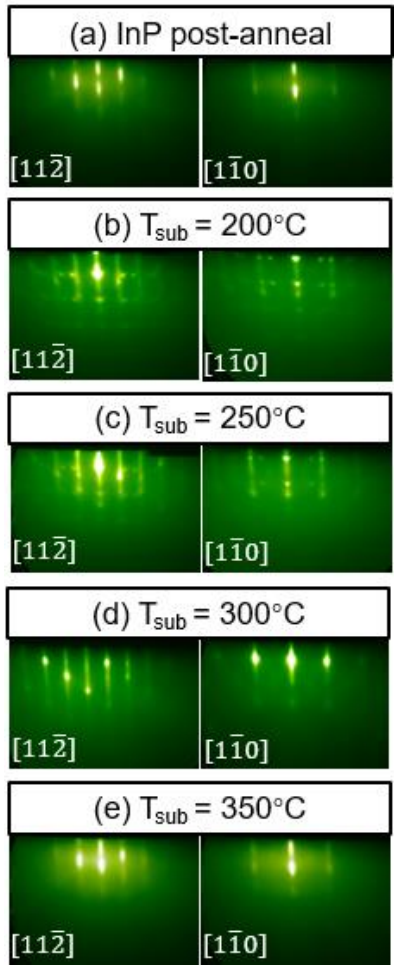

**Figure 1.** (a) RHEED pattern of InP (111)A substrate taken after annealing at 600 °C for 10 minutes in Te environment. (b-e) RHEED patterns taken immediately after the growths of SnTe layers with T$_{sub}$ of (b) 200 °C, (c) 250 °C, (d) 300 °C, and (e) 350 °C.

### B. Effect of Te:Sn flux ratio on the film surface morphology

**Table 1.** Growth parameters of Samples A to D including Sn flux, Te flux, Te:Sn flux ratio, film growth rate. The film surface RMS roughness calculated from a 2 × 2 µm² AFM image for each sample is also listed.

| Sample No. | Sn Flux (cm$^{-2}$/s) | Te Flux (cm$^{-2}$/s) | Te:Sn Flux Ratio | SnTe Growth Rate (Å/s) | RMS roughness (nm) |
|---|---|---|---|---|---|
| A | $3.8 \times 10^{13}$ | $3.8 \times 10^{13}$ | 1 | 0.24 | 6.5 |
| B | $3.8 \times 10^{13}$ | $7.6 \times 10^{13}$ | 2 | 0.24 | 3.5 |
| C | $3.8 \times 10^{13}$ | $1.1 \times 10^{14}$ | 3 | 0.24 | 0.9 |
| D | $3.8 \times 10^{13}$ | $1.9 \times 10^{14}$ | 5 | 0.24 | 6.9 |

Within the optimal substrate temperature window (300-340 °C), the effect of the tellurium to tin (Te:Sn) flux ratio on the surface morphology of the SnTe film was further evaluated. Table 1 summarizes the growth conditions for a series of samples in which the flux ratio was changed while the substrate temperature, growth rate, and film thickness were held constant at 340 °C, 0.24 Å/s, and ~15 nm, respectively. The surface morphology and *in-situ* RHEED patterns of these samples are presented in Figure 2. By comparing the film surface morphologies of Samples A, B and C, it is clear that the film coalescence improves as the Te:Sn flux ratio increases from 1 to 3. The rough surface observed in Samples A and B may be caused by the high desorption rate of tellurium. It was determined previously, that the desorption rate of tellurium exceeds 0.8 Å/s at this substrate temperature (340 °C).[46, 47] Tellurium desorption may lead to a tin-rich growth front, resulting in randomly-shaped SnTe nuclei rather than lateral film coalescence with atomic triangular terraces observed in Figure 2(c). The AFM image for Sample C (Figure 2(c)) with a 3:1 Te:Sn flux ratio shows a coalesced SnTe film without any granular features; its RMS roughness of 0.91 nm is the lowest in this series of samples. However, small triangular voids are still observed in Sample C. The area density of these voids (determined by measuring the number of voids in a larger 10 × 10 µm$^2$ AFM image shown in Figure S2), is ~ 3 × 10$^9$ cm$^{-2}$. In an attempt to reduce the density of voids, the Te:Sn flux ratio was further increased to 5 in Sample D. Unfortunately, this resulted in the surface becoming granular again, with reduced coalescence and an increased RMS roughness (Figure 2(d)). The increased roughness and incomplete coalescence may be attributed to a severely reduced tin adatom diffusion length due to the over-supply of tellurium, resulting in island formation instead of a coalesced layer. In summary, a Te:Sn flux ratio of 3 yields a coalesced SnTe film with an RMS roughness less than 1 nm.

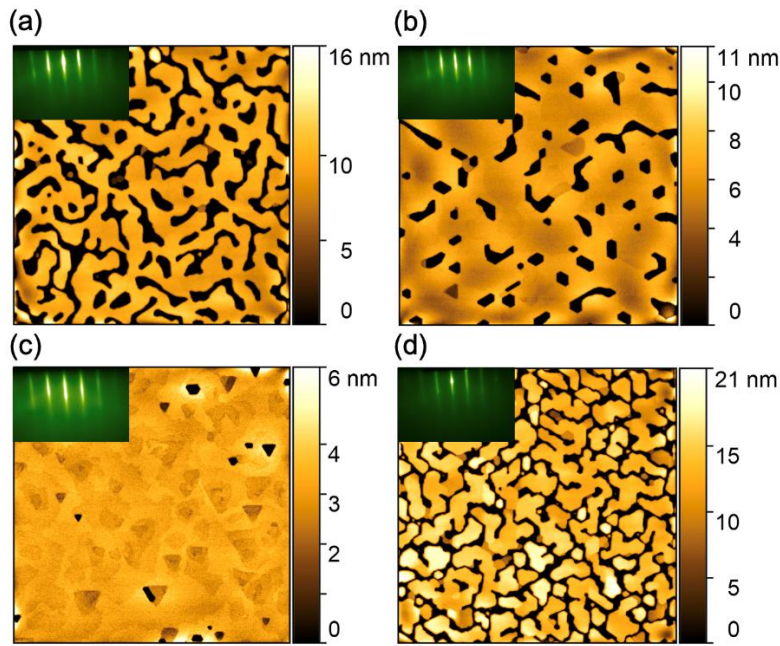

**Figure 2.** AFM images of SnTe films grown with a Te:Sn flux ratios of (a) 1 – sample A, (b) 2 – sample B, (c) 3 – sample C, and (d) 5 – sample D, at a rate of 0.24 Å/s. All AFM scans are 2 × 2 µm$^2$. Each inset shows the corresponding RHEED pattern along the [$11\bar{2}$] direction taken immediately after growth.

## C. Effect of growth rate on film surface morphology and crystallinity

**Table 2.** Summary of the growth parameters of Samples E to G; Sn flux, Te flux, Te:Sn flux ratio, and film growth rate. The RMS film surface roughness measured from a 2 × 2 µm$^2$ AFM image for each sample is also listed.

| Sample No. | Sn Flux (cm$^{-2}$/s) | Te Flux (cm$^{-2}$/s) | Te:Sn Flux Ratio | SnTe Growth Rate (Å/s) | RMS roughness (nm) |
|---|---|---|---|---|---|
| E | 9.5 × 10$^{12}$ | 2.9 × 10$^{13}$ | 3 | 0.06 | no film growth |
| F | 1.9 × 10$^{13}$ | 5.8 × 10$^{13}$ | 3 | 0.12 | 5.6 |
| C | 3.8 × 10$^{13}$ | 1.14 × 10$^{14}$ | 3 | 0.24 | 0.9 |
| G | 7.6 × 10$^{13}$ | 2.3 × 10$^{14}$ | 3 | 0.48 | 0.2 |

The effect of growth rate on the quality of the SnTe thin film was studied next. The substrate temperature was kept at 340 °C, the Te:Sn flux ratio at ~ 3, and the film thickness at 15 nm for Sample E (grown at a rate of 0.06 Å/s), Sample F (grown at a rate of 0.12 Å/s), and Sample G (grown at a rate of 0.48 Å/s). These samples are directly compared to Sample C (discussed in the previous section and grown at a rate of 0.24 Å/s). All growth conditions for this series of samples are listed in detail in Table 2. Figure 3 shows AFM images of the films along with the corresponding RHEED patterns for the four samples. For comparison purposes, Figure 2(c) for Sample C is presented again as Figure 3(c). During the growth of Sample E with the lowest growth rate, the *in situ* RHEED pattern remained largely unchanged from the typical InP RHEED patterns, indicating no SnTe crystallites nucleated. Figure 3(a) therefore shows the bare InP substrate with no SnTe film. The lack of a film is likely due to a combination of the nonzero re-evaporation of SnTe at 340 °C and the slow growth rate. If the rate of arriving Sn adatoms is lower than the re-evaporation rate, no nuclei will be able to form. A growth rate of 0.12 Å/s (Sample F) results in a rough surface with SnTe nano island formation, as shown in Figure 3(b). Compared to Sample C, the surface in Sample F is not coalesced, and the orientations of SnTe nano islands are not well-aligned. This may be also be caused by the nonzero re-evaporation of SnTe at 340 °C or by an imperfect wetting of the substrate by the SnTe film. It is also possible that films grown using these conditions are highly strained, leading to cracking and island formation.

Contrary to the rough and uncoalesced surfaces obtained with a slow growth rate, a smooth SnTe film was obtained with a fast growth rate of 0.48 Å/s (Sample G), as shown in Figure 3(d). The SnTe film of Sample G is similar to Sample C but shows an even further reduced RMS roughness of 0.2 nm compared to Sample C (0.9 nm). In contrast to Sample C, where several triangular voids can be observed, the surface of Sample G is almost free of voids (as shown by the larger-area AFM image in Figure S2 in the supporting information) and shows clear step edges. The step height of ~ 0.6 nm between surface terraces is consistent with the lattice constant of SnTe. The presence of atomic terraces free of bi-layer nucleation sites indicates a step flow growth mode for the SnTe film. The density of the triangular voids in Sample C, as measured by the number of voids over the entire area in Figure S1, is ~$3 \times 10^9$ cm$^{-2}$; the void density is reduced by three orders of magnitude to $5 \times 10^6$ cm$^{-2}$ for Sample G. The improvement in surface morphology with increasing growth rate is somewhat counterintuitive but can be explained by realizing that the higher growth rate suppresses the diffusion of metal (Sn) atoms and limits 3D island formation. Similar phenomena are observed in III-V semiconductors such as InAs.[48] It is expected that further increasing the growth rate would lead to further improvements in surface morphology. However, the maximum growth rate in this study was kept at 0.48 Å/s (Sample G) due to the physical limitations of effusion cell operations for both elements Sn and Te.

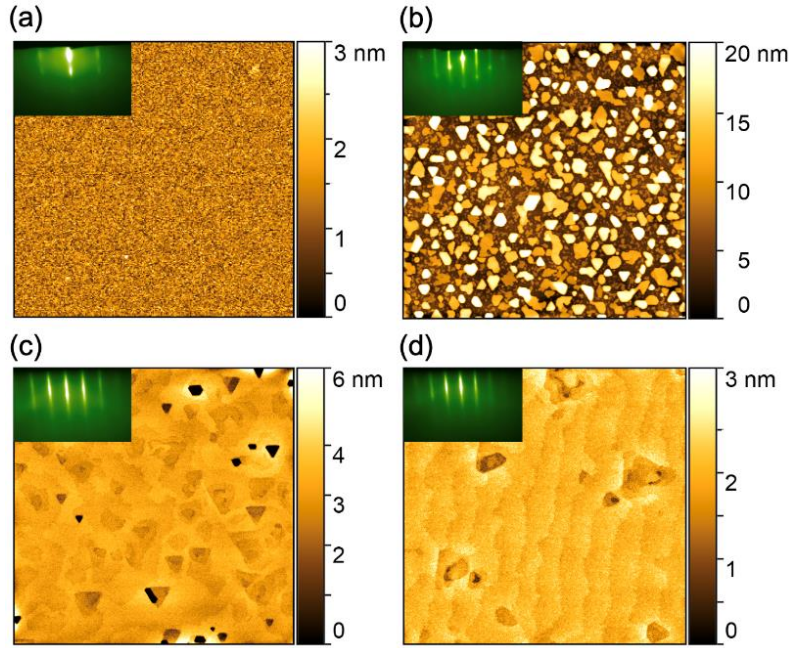

**Figure 3.** $2 \times 2$ μm$^2$ AFM scans of Sample E (a), F (b), C (c), and G (d), grown at a rate of 0.06 Å/s, 0.12 Å/s, 0.24 Å/s, 0.48 Å/s, respectively. Panel (c) is the same as Figure 2(c). Each inset shows the corresponding RHEED pattern along the [11$\bar{2}$] direction taken immediately after growth.

Figure 4 shows the XRD 2θ-ω scans for Samples F, C, and G, respectively (sample E is not shown as no film formed). Single phase SnTe (111) films were observed for all three samples, evidenced by the strong SnTe (222) diffraction peak at ~ 50°. An additional peak at ~ 25° was observed in both Samples C and G, which was identified as the SnTe (111) diffraction peak. The exact position of the SnTe (111) diffraction peak was, however, impossible to extract due to its overlap with the InP (111) substrate diffraction peak at 26°. Nonetheless, by applying Bragg's Law using the (222) peak,[49] the out of plane lattice constant $a_0$ of cubic SnTe was calculated to 6.31 Å, which is in good agreement with previously reported values.[32, 33, 50] It is worth noting that interference fringes due to Pendellösung oscillations can be clearly observed surrounding both the SnTe (111) and (222) diffraction peaks in Sample G, indicative of an abrupt and sharp interface between the SnTe film and the InP substrate, in addition to the atomically flat SnTe film surface.[51] It is thus evident that a high growth rate leads to a flatter SnTe layer and benefits greatly its long-range crystallographic order.

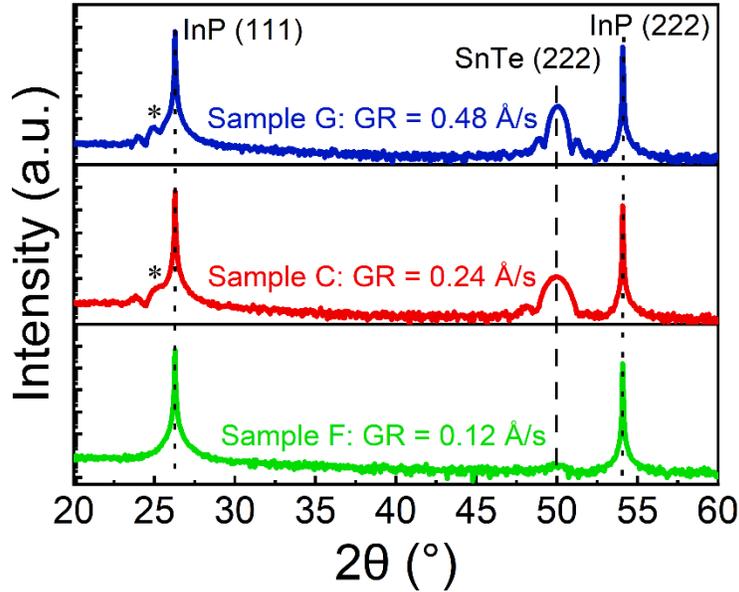

**Figure 4.** XRD 2θ-ω scans for Samples F - green, C - red, and G - blue grown at a growth rate of 0.12 Å/s, 0.24 Å/s, and 0.48 Å/s, respectively. The vertical short-dashed lines highlight the InP (111) and (222) substrate diffraction peaks and the vertical long-dashed line indicates the position of the SnTe (222) peaks. The peak or bump marked by "*" denotes the approximate position of the SnTe (111) peaks.

D. **Optimized SnTe thin films and their structural properties**

High resolution XRD scans were performed to investigate the crystalline quality of Sample G – grown under optimized conditions as determined above (340 °C, Te:Sn flux ratio of 3, and growth rate of 0.48 Å/s) – in detail. Figures 5(a) and (b) show rocking curves around the SnTe (222) and

(220) diffraction peaks. Both curves have a sharp and single-peak Gaussian-like distribution. By applying a Pseudo-Voigt fitting profile,[52] the FWHM extracted from the SnTe (222) rocking curve is 0.09°, while that of the SnTe (220) is 0.42°. These values are the lowest reported for (111)-oriented SnTe films to date, which confirm the exceptional crystalline quality in these films. Equation (1) from Ref. [53] was used to estimate the threading dislocation density in the SnTe film where a Gaussian profile was assumed in the random distributions of threading dislocation[53-55]:

$$D = \beta^2/(2\pi \ln 2 b^2) \approx \beta^2/(4.36 b^2) \quad (1)$$

where $D$ is the dislocation density, $\beta$ is the FWHM of the XRD rocking curve, and $b$ is the length of the Burgers vector. Here, $b$ was assumed to be the out-of-plane lattice constant (6.31 Å) for both screw-type dislocations and edge-type dislocations. Applying equation (1), the density of screw-type (and mixed-type) dislocations was found to be $1.6 \times 10^8$ cm$^{-2}$, while the density of edge-type (and mixed-type) dislocations was about ~ $3.1 \times 10^9$ cm$^{-2}$. The total dislocation density can thus be estimated to be ~ $3.3 \times 10^9$ cm$^{-2}$. Considering the thickness of these films of only ~ 15 nm paired with > 7% lattice mismatch to the substrate, this exceptionally low value highlights the enormous potential for using these films as buffer layers for other tellurides. Furthermore, it is expected that both the XRD rocking curve FWHM and the dislocation density can be further reduced by simply increasing the SnTe film thickness.

Finally, the XRD in-plane φ-scan around the SnTe 220 reflection is shown in Figure 5(c). An in-plane rotational φ-scan is an effective method for measuring the density of rotational twin domains in the film.[56] The φ-scan for InP along the 220 reflection is presented as a reference. For the InP scan (shown in the blue dashed line), three peaks can be observed, corresponding to the three-fold in-plane rotational symmetry of the (111)-oriented substrate. For the SnTe film, mimicking the InP substrate, an identical set of only three peaks separated by 120° is observed, indicating that the film is free of twin defects which act as scattering centers in electronic devices and nonradiative recombination pathways in optical devices. In the past, twin defects have been observed in SnTe films grown on CdTe (111) buffer layers,[57] which was attributed to the presence of domains with different stacking orders.[58] In the current study, such twin domains have been eliminated, further confirming the excellent crystallinity in the as-grown optimized SnTe films.

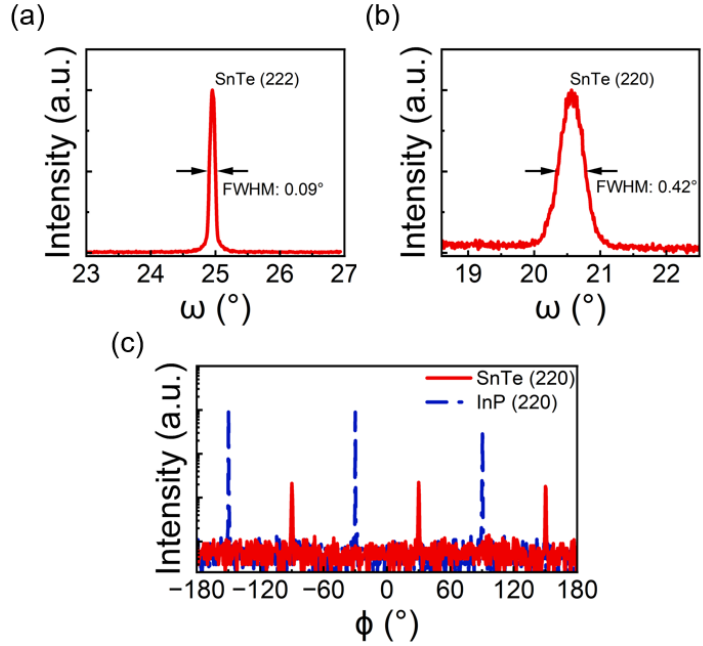

**Figure 5.** XRD rocking curves of SnTe along the (a) (222) and (b) (220) directions. (c) In-plane φ-scan of the SnTe (220) diffraction peak. The φ-scan for the InP substrate (220) diffraction peak is presented in the blue dashed line for reference.

To evaluate the strain state and degree of relaxation of optimized SnTe films, i.e., Sample G, reciprocal space mapping (RSM) was employed. Figures 6(a) and (b) show the RSM spectra around the symmetrical [222] and asymmetrical [224] directions. Both reciprocal space maps depict two clear peaks, one corresponding to the InP substrate and one corresponding to the SnTe film. The well-defined centers in the mosaic spread of both the SnTe (222) and SnTe (224) spots suggest that the as-grown SnTe epilayer has a high degree of crystalline alignment and structural coherence. In Figure 6(a), the Bragg peaks of SnTe (222) and InP (222) are both on the $Q_x = 0$ line, confirming that the SnTe (111) planes are fully aligned with the InP substrate with no miscut or tilt. In Figure 6(b), the relaxation of the SnTe film relative to the InP substrate is investigated. In a fully relaxed cubic structure, a theoretical angle of 19.471° is expected between the [111] and [224]. The angle between the mosaic center of the InP (224) reflection and the [111] direction was calculated to be 19.466° from the RSM in Figure 6(b), which agrees well with the theoretical value. A similar good agreement to the theoretical value was calculated from the RSM for the angle between the SnTe (224) mosaic center and the [111] direction of 19.461°, suggesting the current 15 nm epitaxial SnTe film is fully relaxed despite the > 7% lattice mismatch between film and substrate.

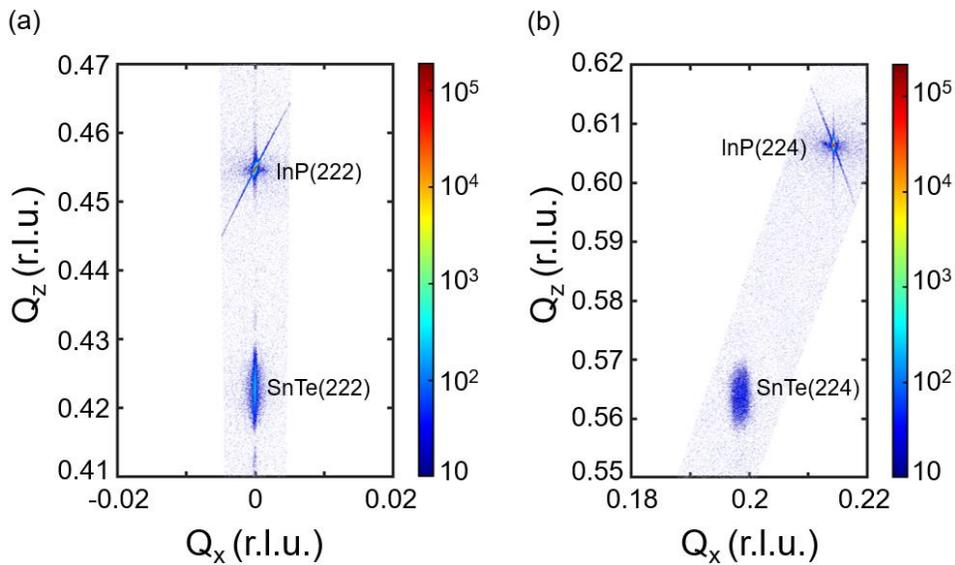

**Figure 6.** Reciprocal space maps of XRD peaks around the (a) InP (222) and (b) InP (224) diffractions. Both SnTe (222) and SnTe (224) mosaics are clearly observed and labeled in both figures, respectively.

Considering the relatively low dislocation density in the as-grown SnTe layer as evidenced from the XRD rocking curves, one would expect a SnTe layer of ~15 nm to be partially strained to the substrate, since the critical thickness of SnTe on InP is ~4.5 nm. However, as discussed above,

RSM indicates that the present 15 nm-thick SnTe film is fully relaxed. To elucidate the relaxation mechanism, STEM imaging at the SnTe/InP interface was performed. High-angle annular dark field (HAADF) STEM image of the SnTe/InP heterointerface along the [110] zone axis is shown in Figure 7(a). A highly crystalline SnTe layer can be observed, with brighter contrast as compared to the InP substrate region. A hazy atomically thin region presents at the SnTe/InP heterointerface, which generally indicates the presence of dislocations.[39] To clarify the positions of the dislocations, Fast Fourier Transform (FFT) analysis was performed using the image in Figure 7(a). After using a masking filter of the FFT image around the [-111] spots, we then perform the inverse FFT analysis, following the technique of Ref.[43], which is shown in Figure 7(b). This analysis clearly highlights the interfacial misfit dislocations (IMFs) that are uniformly spaced. The spacing between each misfit dislocation is ~ 5.2 nm, which corresponds to 13 SnTe lattice sites and 14 InP lattice sites. To characterize the origin of the misfit dislocations, a right-hand Burger's circuit was constructed starting in the InP substrate, as demonstrated in Figure 7(c). This resulted in a Burgers vector (***b***) with a length of 1 InP lattice site that lies completely in the plane of the SnTe/InP (111) heterointerface, which indicates that the Burgers vector is of the $\frac{a}{2}\langle 110 \rangle$ type.[43] In III/V on III/V heteroepitaxy (e.g. GaSb on GaAs), IMF arrays have been previously observed to serve as the main mechanism for rapid strain relaxation yielding ultralow threading dislocation densities in the film.[40, 42, 43] Unlike previous observations, this work demonstrates an IMF array in a system with heterogeneous crystal structures, i.e., rock-salt epitaxial layer on a zinc-blende substrate, and furthermore in a system with differing cations and anions in the film and substrate, i.e. a IV-VI semiconductor film grown on a III-V substrate. Finally, the IMF array was observed for a film and substrate in the (111) orientation, which was only recently demonstrated in III-V heteroepitaxy.[43]

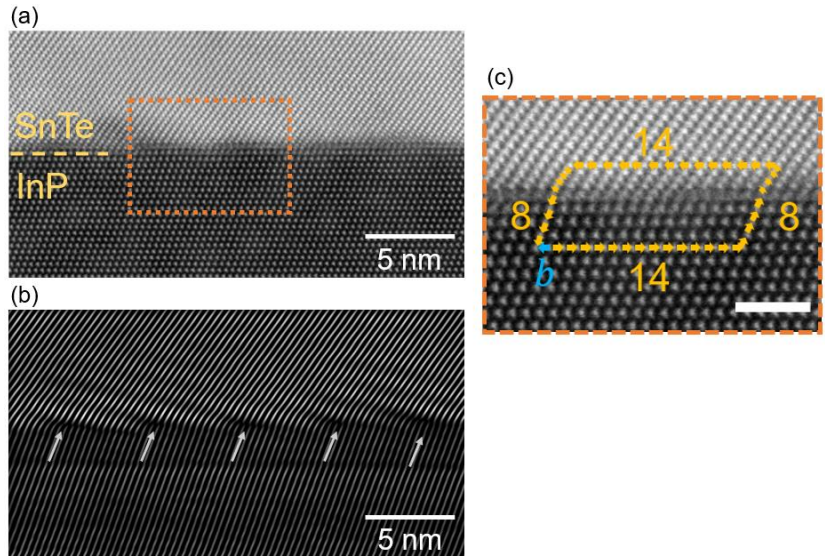

**Figure 7.** (a) HAADF-STEM image of SnTe film on InP (111)A. The dashed line indicates the interface between the SnTe film and the InP substrate. (b) Inverse FFT analysis of (a) using the [-111] spot. Each arrow highlights an interfacial misfit dislocation at the heterointerface. (c) A constructed Burgers circuit in the boxed region (marked in orange) of (a). The resulting Burgers vector is denoted in blue. The scale bar is 2 nm.

## 4. Summary

In summary, this work presents detailed investigations of MBE-grown SnTe epitaxial layers and discusses how the growth conditions impact surface morphology and crystalline quality. An increase of substrate temperature from 200 °C to 300-340 °C results in a transition from polycrystalline to single crystalline SnTe formation. An optimized Te:Sn flux ratio is found to lead to atomically smooth SnTe layers with triangular void formation. As the growth rate of the SnTe film increases, the surface morphology is substantially improved, as the density of triangular voids is reduced by three orders of magnitude. With these optimized growth parameters, large-area and high-quality SnTe (111) layers with a RMS roughness as low as 0.2 nm and an XRD rocking curve FWHM as low as 0.09°, both of which are state-of-the-art values, is obtained. The film is free of rotational twin domains, and the strain is fully relaxed despite the > 7 % lattice mismatch between film and substrate. Surprisingly, this work uncovers that the strain is relieved through the formation of a periodic interfacial misfit array at the SnTe/InP heterointerface, where each misfit dislocation is evenly spaced at 13 SnTe and 14 InP lattice sites. This highly efficient IMF array relaxes the lattice strain rapidly and leads to the observed high quality of the SnTe film. This is the first demonstration of the formation of an IMF array in a rock-salt on zinc-blende material system, as well as the first instance of such between an IV-VI epitaxial layer on a III-V substrate. This result highlights that IMF arrays can be used to relax strain in a much wider range of heteroepitaxial systems than previously realized. These important observations thus facilitate the heterointegration of an important IV-VI semiconductor on a III-V substrate, unlocking the potential of creating high performance multi-functional devices. The smooth, relaxed, single crystalline SnTe films could also be used for infrared detectors, infrared plasmonics, and as a model system to understand the behavior of topological crystalline insulators when integrated with an InP back gate.

ACKNOWLEDGMENT

This research was conducted at the Pennsylvania State University Two-Dimensional Crystal Consortium – Materials Innovation Platform which is supported by NSF cooperative agreement DMR-2039351. The authors appreciate the use of the Penn State Materials Characterization Lab.

AUTHOR INFORMATION

**Corresponding Author**

Qihua Zhang – qzz5173@psu.edu

Stephanie Law – sal6149@psu.edu

**Data availability statement**

All data of this study is available under the following link for the review process, which will be converted into an open-access link to the data with separate DOI in ScholarSphere upon publication: https://data.2dccmip.org/aN3h8cyZuPAP.

**Supplementary Materials**

# Growth Conditions and Interfacial Misfit Array in SnTe (111) films Grown on InP (111)A Substrates by Molecular Beam Epitaxy

*Qihua Zhang[1], Maria Hilse[1,2,3], Wesley Auker[2], Jennifer Gray[2], Stephanie Law[1,2,3,4]*

[1]Two Dimensional Crystal Consortium Materials Innovation Platform, The Pennsylvania State University, University Park, PA 16802, USA

[2]Materials Research Institute, The Pennsylvania State University, University Park, Pennsylvania 16802 USA

[3]Department of Materials Science and Engineering, The Pennsylvania State University, University Park, Pennsylvania 16802 USA

[4]Institute of Energy and the Environment, The Pennsylvania State University, University Park, PA 16802 USA

Figure S1 shows the x-ray reflectivity (XRR) measurements of SnTe epitaxial layers (Sample G) grown on InP (111)A substrate. The fitting analysis (shown in red dashed line) is obtained using Advanced Material Analysis and Simulation Software (AMASS) which is included in the Malvern PANalytical 4-circle X'Pert$^3$ system. The as-fitted SnTe layer thickness ($t_{SnTe,\ fit}$) is ~14.6 nm, which agrees well with the target thickness (15 nm) for the growth.

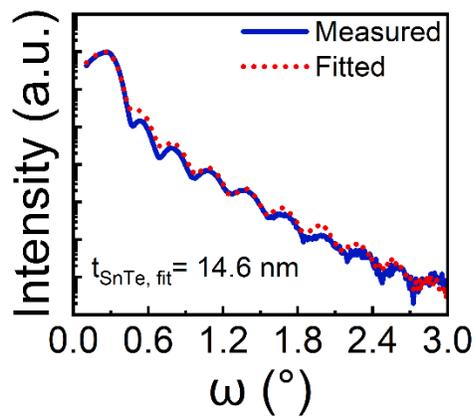

**Figure S1.** Measured X-ray reflectivity (XRR) data (in blue solid line) and the simulated fitting result (red dashed line) of SnTe (111) layer grown on InP (111)A substrate (Sample G).

Figure S2 displays the large-area (scan area: 10 × 10 µm²) AFM scans for Sample C (a) and Sample G (b). The RMS roughness for (a) and (b) are 1.65 nm and 0.33 nm, respectively. The density of voids was measured by counting the number of voids in the entire scan area, which is estimated to be ~3 × 10$^9$ cm$^{-2}$ for Sample C, and 5 × 10$^6$ cm$^{-2}$ for Sample G.

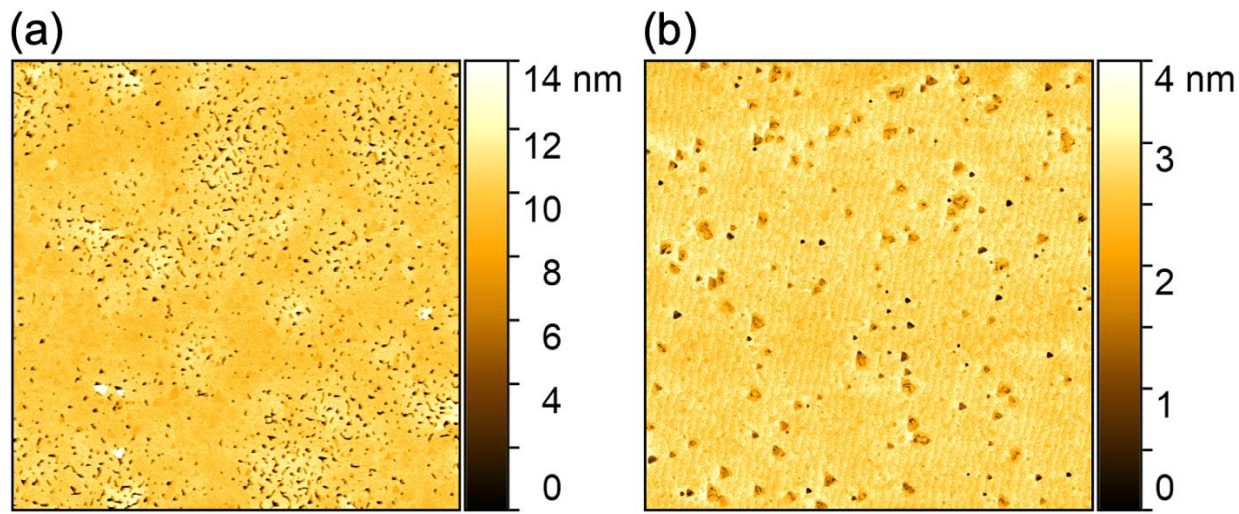

**Figure S2.** Large-area AFM images for Sample C (a) and G (b). The scan area is 10 × 10 µm² for both samples.